\documentclass[aps,prb,showpacs,twocolumn,nofootinbib,superscriptaddress]{revtex4}
\usepackage{amsmath,graphicx,epsfig,amssymb,subfigure,times,dsfont}
\usepackage[usenames]{color}

\newcommand{\ket}[1] {| #1 \rangle}
\newcommand{\bra}[1] {\langle #1 |}
\newcommand{\braket}[2] {\langle #1 | #2 \rangle}
\newcommand{\proj}[1] {| #1 \rangle\langle #1 |}
\newcommand{\tr} {\mbox{tr }}

\begin{document}

\title{Perfect Sampling with Unitary Tensor Networks}

\author{Andrew~J.~Ferris}
\affiliation{The University of Queensland, School of Mathematics and Physics, Queensland 4072, Australia}
\affiliation{D\'epartement de Physique, Universit\'e de Sherbrooke, Qu\'ebec, J1K 2R1, Canada}
\author{Guifre~Vidal}
\affiliation{The University of Queensland, School of Mathematics and Physics, Queensland 4072, Australia}
\affiliation{Perimeter Institute for Theoretical Physics, Waterloo, Ontario, N2L 2Y5, Canada}

\date{\today}
\begin{abstract}
Tensor network states are powerful variational ans\"atze for many-body ground states of quantum lattice models. The use of Monte Carlo sampling techniques in tensor network approaches significantly reduces the cost of tensor contractions, potentially leading to a substantial increase in computational efficiency. Previous proposals are based on a Markov chain Monte Carlo scheme generated by locally updating configurations and, as such, must deal with equilibration and autocorrelation times, which result in a reduction of efficiency. Here we propose perfect sampling schemes, with vanishing equilibration and autocorrelation times, for unitary tensor networks -- namely tensor networks based on efficiently contractible, unitary quantum circuits, such as unitary versions of the matrix product state (MPS) and tree tensor network (TTN), and the multi-scale entanglement renormalization ansatz (MERA). Configurations are directly sampled according to their probabilities in the wave-function, without resorting to a Markov chain process. We consider both complete sampling, involving all the relevant sites of the system, as well as incomplete sampling, which only involves a subset of those sites, and which can result in a dramatic (basis-dependent) reduction of sampling error.
\end{abstract}


\maketitle

\section{Introduction}
\label{sec:intro}

To the computational physicist interested in one-dimensional quantum lattice models,
the density matrix renormalization group (DMRG) \cite{White92,White93} is a dream come true. It provides an essentially unbiased, extremely accurate variational approach to ground state properties of a large class of local Hamiltonians in one dimensional lattices. DMRG operates by approximating the ground state of the system with a matrix product state (MPS) \cite{Fannes92,Ostlund95,Vidal03,Perez-Garcia07}, which is a simple tensor network with tensors connected according to a one-dimensional array. In recent years, the success and broad applicability of DMRG has been understood to follow from (i) the existence of a characteristic, universal pattern of entanglement common to most ground states in one spatial dimension; and (ii) the ability of the MPS to reproduce this universal pattern of entanglement, thanks to having its tensors connected into a one-dimensional geometry.

The above insight has since then guided the development of new tensor network approaches that aim to repeat, in other geometries or physical regimes of interest, the unprecedented success of DMRG \cite{White92, White93,Schollwoeck05,Schollwoeck11} in one dimension. The recipe is quite simple: first, identify a pattern of entanglement common to a large class of ground states; then, connect tensors so that they can reproduce this pattern, and use the resulting tensor network as a variational ansatz. In this way the multi-scale, layered pattern of entanglement observed in ground states near a continuous quantum phase transition motivated the proposal of the multi-scale entanglement renormalization ansatz (MERA) \cite{Vidal07,Vidal10} to address quantum critical phenomena. Similarly, the characteristic spatial pattern of entanglement in the ground states in two and higher dimensions motivated higher-dimensional generalizations of both the MPS (known as projected entangled-pair states, PEPS\cite{Verstraete04, Sierra98, Nishino98,Murg07,Jordan08,Gu08,Jiang08}) and the MERA \cite{Evenbly10,Evenbly10b,Cincio08,Evenbly09b}.

The cost of simulating a lattice of $L$ sites with any of the above tensor networks is roughly proportional to $L$, which underlies the efficiency of the approaches\cite{reduction}. Importantly, however, this cost also grows as $O(\chi^p)$, that is as a power $p$ of the dimension $\chi$ of the indices connecting the tensors into a network. On the one hand, this bond dimension $\chi$ determines the size of the tensors and therefore the number of variational parameters contained in the tensor network ansatz. On the other, $\chi$ is also a measure of how much entanglement the tensor network can carry. It then follows that the cost of simulations increases with the amount of entanglement in the ground state of the system. Entanglement is indeed the key factor limiting the range of applicability of tensor network approaches.

More specifically, for an MPS, a small power $p$, namely $p_{\mbox{ \tiny MPS}} = 3$, implies that very large values of $\chi$ (of up to a few thousands) can be considered even with a high-end desktop computer. Correspondingly, DMRG can address one-dimensional systems with robustly entangled ground states. In contrast, the cost of two dimensional simulations with PEPS and MERA scales with a much larger power $p$ of $\chi$, e.g. $p_{\mbox{ \tiny PEPS}} = 12$ in Ref. \onlinecite{Murg07} and $p_{\mbox{ \tiny MERA}} = 16$ in Ref. \onlinecite{Evenbly09b}, and this considerably reduces the affordable values of $\chi$. In other words, PEPS and MERA calculations have so far been restricted to systems with relatively small amounts of ground state entanglement. A major present challenge for these approaches is to obtain more efficient tensor contraction schemes that could lower their cost.

A possible route to reducing the scaling of computational cost with $\chi$ in tensor network algorithms is by using Monte Carlo sampling techniques, as proposed in Refs. \onlinecite{Schuch08, Sandvik07, White09}. As reviewed in the next section, the cost of manipulating the tensor network (for a single sample) is reduced to $O(\chi^{q})$, where $q$ is significantly smaller than $p$ (typically of the order of $p/2$). The proposals in Refs. \onlinecite{Schuch08, Sandvik07} are best suited for tensor networks, such as MPS and PEPS, where the coefficients in the tensors are unconstrained. However, in the MERA, as well as in other unitary tensor networks such as unitary versions of MPS (uMPS) and of tree tensor network\cite{Shi06,Tagliacozzo09} (uTTN), tensors are subject to unitary constraints.

The purpose of this paper is to address the use of Monte Carlo sampling in the context of unitary tensor networks, including uMPS, uTTN and MERA. [Notice that this excludes tensor networks such as a periodic MPS or PEPS, which cannot be generically re-expressed as a unitary tensor network]. An important difference with respect to Refs. \onlinecite{Schuch08, Sandvik07} is that in a unitary tensor network, sampling is performed on an effective lattice corresponding to the past causal cone of the local operator whose expectation value is being computed. This means that sampling typically occurs over some reduced number of sites (less than the system size $L$). A second difference is that in unitary tensor networks there is no need to use a Markov chain Monte Carlo scheme. Indeed, our main result is the proposal and benchmark of perfect sampling schemes for unitary tensor networks, by means of which one can obtain completely uncorrelated samples directly according to the correct probability. Therefore, one can sample without incurring additional computational costs due to equilibration and autocorrelations times. This is particularly of interest near a quantum phase transition, where equilibration and autocorrelation times diverge with system size $L$. We consider both complete (perfect) sampling and incomplete (perfect) sampling schemes. In the former, the indices for all sites of the effective lattice are sampled. In the latter, only the indices of a subset of sites is sampled, while the indices of the rest of sites are contracted exactly, with an insignificant or minor increase of computational cost as far as the scaling $O(\chi^q)$ is concerned. Importantly, the statistical variance (due to sampling) of an expectation value obtained with incomplete sampling can decrease dramatically with a proper chose of sampling basis, as illustrated in Fig. \ref{fig:partialMPS3} with a drop of $10^{-7}$ in error.

The paper is organized in sections as follows. First, in section \ref{sec:background} we briefly review the use of Monte Carlo sampling techniques to evaluate the expectation value of local operators in context of tensor networks, and introduce the notions of complete and incomplete sampling. Then in section \ref{sec:unitary} we explain how the proposals of Refs. \onlinecite{Schuch08, Sandvik07} can be adapted to the case of a unitary tensor network by sampling within the past causal cone of the local operator. In section \ref{sec:perfect} we propose a complete perfect sampling scheme for unitary tensor networks. Its performance is demonstrated for a uMPS with the quantum Ising chain at criticality. In section \ref{sec:partial} we then present an incomplete perfect sampling scheme. We discuss computational costs in section \ref{sec:costs}. The conclusions in Section \ref{sec:conclusion} and an Appendix analyzing the variance in different schemes close the paper.

We emphasize that this paper is only concerned with the evaluation of local expectation values from a unitary tensor network. That is, here we assume that the unitary tensor network has already been optimized and focus on how to extract information from it. The optimization of unitary tensor networks using variational Monte Carlo is discussed in Ref. \onlinecite{Ferris11}.

\section{Background material: sampling in tensor network algorithms}
\label{sec:background}

Let us start by introducing our notation and by reviewing some basic concepts.

\subsection{Exact contraction versus sampling}


Let $\mathcal{L}$ be a lattice made of $L$ sites, with vector space $\mathbb{V}_{\mathcal{L}} \equiv \otimes_{i=1}^L \mathbb{V}$, where $\mathbb{V}$ is the $d$-dimensional vector space of one site. Let $\ket{\Psi}\in \mathbb{V}_{\mathcal{L}}$ denote the wave-function encoded in the tensor network and let $\hat{A}$ be a local operator on $\mathbb{V}_{\mathcal{L}}$. An important task in tensor network algorithms is to compute the expectation value $\bra{\Psi}\hat{A}\ket{\Psi}$, which can be expressed as
\begin{equation}
	\bra{\Psi} \hat{A} \ket{\Psi} = \sum_{\mathbf{s}\in\mathcal{S}} \braket{\Psi}{\mathbf{s}}\bra{\mathbf{s}}\hat{A}\ket{\Psi},
	\label{eq:expect}
\end{equation}
where $\ket{\mathbf{s}} \equiv \ket{s_1}\otimes \ket{s_2} \otimes \cdots \otimes \ket{s_L}$ denotes a product state of the $L$ sites of the lattice, with $s_i=1,2,\cdots, d$ labelling the elements of an orthonormal basis $\{ \ket{s_i}\}$ on site $i$, $i=1,2, \cdots, L$. Here, $\mathcal{S}$ is the set of all $d^L$ possible \textit{configurations} $\mathbf{s}=(s_1, s_2, \cdots, s_L)$ of the system. The expectation value of Eq.~(\ref{eq:expect}) can be obtained exactly by contracting the corresponding tensor network. However, a large computational cost motivates the search for an alternative approach based on sampling.

\begin{figure}[t]
  \begin{centering}
    \includegraphics[width=8.5cm]{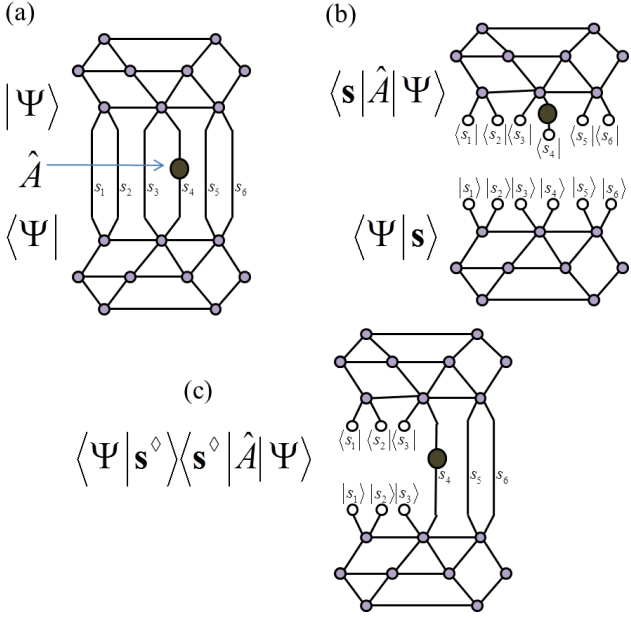}
    \caption{
    (Color online) Contraction of a tensor network. (a) Tensor network corresponding to the expectation value $\bra{\Psi}\hat{A}\ket{\Psi}$, with a sum over (or exact contraction of) indices $s_1, s_2, \cdots, s_6$ (exact contraction). Contracting this tensor network has a cost that scales as $O(\chi^p)$ with the bond index $\chi$, for some power $p$.
    (b) Tensor networks corresponding to $\braket{\Psi}{\mathbf{s}} \bra{\mathbf{s}} \hat{A} \ket{\Psi}$ for a given configuration $\mathbf{s}$, corresponding to a single sample. The cost of contracting these two networks scales as $O(\chi^q)$ with the bond index $\chi$, where power $q$ is smaller than power $p$.
    (c) Tensor network corresponding to $\braket{\Psi}{\mathbf{s}^{\diamond}}\bra{\mathbf{s}^{\diamond}} \hat{A} \ket{\Psi}$ for a given incomplete configuration $\mathbf{s}^{\diamond} \equiv (s_1, s_2, s_3)$ (these three indices are being sampled), where in addition there is a sum over (or exact contraction of) indices $s_4, s_5$ and $s_6$. The cost of contracting this tensor network scales as $O(\chi^{q'})$, with $q'$ somewhere between $q$ and $p$.
    \label{fig:sampling}}
  \end{centering}
\end{figure}


In preparation for an approximate evaluation of the expectation value  $\bra{\Psi} \hat{A} \ket{\Psi}$, let us first introduce the probability $Q(\mathbf{s})\equiv |\braket{\mathbf{s}}{\Psi}|^2$ of projecting state $\ket{\Psi}$ into the product state $\ket{\mathbf{s}}$, and the estimator $A(s) \equiv \bra{\mathbf{s}} \hat{A} \ket{\Psi} / \braket{\mathbf{s}}{\Psi}$, and rewrite Eq.~(\ref{eq:expect}) as
\begin{equation}
 	\bra{\Psi} \hat{A} \ket{\Psi} =\sum_{\mathbf{s}\in\mathcal{S}} Q(\mathbf{s}) A(\mathbf{s}).
\label{eq:average}
\end{equation}
This expression emphasizes that $\bra{\Psi}\hat{A}\ket{\Psi}$ can be regarded as a probabilistic average of estimator $A(\mathbf{s})$ according to the probabilities $Q(\mathbf{s})$, where $Q(\mathbf{s})\geq 0$, $\sum_{\mathbf{s}\in \mathcal{S}} Q(\mathbf{s})=1$.


Let us replace the sum over the set $\mathcal{S}$ of all $|\mathcal{S}|=d^L$ configurations $\mathbf{s}$ with a sum over some subset $\tilde{\mathcal{S}} \subseteq \mathcal{S}$ containing $N \equiv |\tilde{\mathcal{S}}|$ configurations $\mathbf{s}$, where $N < d^L$, that is
\begin{equation}
	\bra{\Psi} \hat{A} \ket{\Psi} \approx  \frac{1}{Z}\sum_{\mathbf{s}\in\tilde{\mathcal{S}}} Q(\mathbf{s}) A(\mathbf{s}),
\label{eq:sampling}
\end{equation}
where $Z \equiv \sum_{\mathbf{s} \in \tilde{\mathcal{S}}} Q(\mathbf{s})$ is a normalization factor. Eq.~(\ref{eq:sampling}) states that an approximate evaluation of $\bra{\Psi} \hat{A} \ket{\Psi}$ is obtained by considering a probabilistic sum over $N$ configurations $\mathbf{s}$. If the $N$ configurations in $\tilde{\mathcal{S}}$ have been randomly chosen from $\mathcal{S}$ according to the probability $Q(\mathbf{s})$, then importance sampling allow us to replace the previous expression with
\begin{equation}
	\bra{\Psi} \hat{A} \ket{\Psi} \approx \frac{1}{N}\sum_{\mathbf{s}\in\tilde{\mathcal{S}}} A(\mathbf{s}).
	\label{eq:importance}
\end{equation}


Equation~(\ref{eq:importance}) estimates $\bra{\Psi} \hat{A} \ket{\Psi}$ by means of $N$ independent samples of a random variable $\left( A(\mathbf{s}),Q(\mathbf{s})\right)$. By construction, the mean $\bar{A}$ of this random variable,
\begin{equation}
	\bar{A} \equiv \sum_{\mathbf{s}\in \mathcal{S}} Q(\mathbf{s})A(\mathbf{s}),
\end{equation}
is given by the expectation value $\bra{\Psi} \hat{A} \ket{\Psi}$ of operator $\hat{A}$, see Eq.~(\ref{eq:average}). Notice that, in addition, its variance $\sigma_{A}^2$, defined by
\begin{eqnarray}
	\sigma_{A}^2 &\equiv& \sum_{\mathbf{s}} Q(\mathbf{s}) |A(\mathbf{s})-\bar{A}|^2 \\
	&=& \sum_{\mathbf{s}} Q(\mathbf{s}) |A(\mathbf{s})|^2-|\bar{A}|^2,
\end{eqnarray}	
also equals the variance $\sigma_{\hat{A}}^2$ of operator $\hat{A}$,
\begin{eqnarray}
\sigma_{\hat{A}}^2	&\equiv&  \bra{\Psi} \left( |\hat{A}-\bra{\Psi}\hat{A}\ket{\Psi}|^2 \right) \ket{\Psi} \\
&=& \bra{\Psi} (|\hat{A}|^2) \ket{\Psi}-|\bra{\Psi} \hat{A} \ket{\Psi}|^2,
\label{eq:varianceAhat}
\end{eqnarray}
that is $\sigma^2_A = \sigma_{\hat{A}}^2$, see Appendix. It follows that the error $\epsilon_A(N)$ in the approximation of Eq.~(\ref{eq:importance}), as measured by the standard deviation $\sigma_{A}/\sqrt{N}$ of $N$ independent samples, scales with $N$ as
\begin{equation}
\epsilon_{A}(N) \approx \sqrt{\frac{ \sigma_{\hat{A}}^2}{N}}.
\label{eq:error}
\end{equation}

Let us analyze in which sense the above Monte Carlo sampling strategy could be of interest.
The cost (i.e. computational time) of an exact contraction, Eq.~(\ref{eq:expect}), scales as $O(\chi^p)$ with the bond dimension $\chi$. On the other hand, notice that for each specific configuration $\mathbf{s}$, the contribution $\braket{\Psi}{\mathbf{s}}\bra{\mathbf{s}}\hat{A}\ket{\Psi}$ to $\bra{\Psi} \hat{A} \ket{\Psi}$ consists of two tensor networks, namely one for $\braket{\Psi}{\mathbf{s}}$ and another for $\bra{\mathbf{s}}\hat{A}\ket{\Psi}$, whose contraction can be accomplished with a cost $O(\chi^q)$, for some $q<p$, see Fig. \ref{fig:sampling}. [This is also the cost of computing $Q(\mathbf{s})$ and $A(\mathbf{s})$ in Eq.~(\ref{eq:average})]. If the number of samples required to obtain an acceptably small error $\epsilon_{A}(N)$ is $N \approx O(\chi^{q'})$, the use of sampling incurs a computational cost of $O(\chi^{q+q'})$ instead of $O(\chi^p)$. We conclude that if $q+q' < p$, then (for large $\chi$) the sampling strategy will have a lower computational cost than the exact contraction.

\subsection{Combining exact contraction with sampling: \\ Incomplete sampling}

More generally, one can consider a hybrid strategy which combines exact contraction and sampling. This is accomplished by sampling over only a subset of the $L$ indices corresponding to the $L$ sites of lattice $\mathcal{L}$, while performing an exact contraction on the remaining sites. For instance, Fig. \ref{fig:sampling}(c) considers a lattice $\mathcal{L}$ made of $L=6$ sites where the first three sites are being sampled, with configuration $(s_1,s_2,s_3)$, whereas the remaining three of sites are being addressed with an exact contraction.

If we denote by $\mathbf{s}^{\diamond} \in \mathcal{S}^{\diamond}$ a configuration of the $L^{\diamond}$ indices to be sampled ($L^{\diamond}< L$), then Eq. \ref{eq:expect} is replaced with
\begin{equation}
	\bra{\Psi} \hat{A} \ket{\Psi} = \sum_{\mathbf{s}^{\diamond}\in\mathcal{S}^{\diamond}} \braket{\Psi}{\mathbf{s}^{\diamond}}\bra{\mathbf{s}^{\diamond}}\hat{A}\ket{\Psi},
	\label{eq:expect_dia}
\end{equation}
We can again rewrite Eq.~(\ref{eq:expect_dia}) as a probabilistic sum of an estimator $A^{\diamond}(\mathbf{s}^{\diamond}) \equiv  \braket{\Psi}{\mathbf{s}^{\diamond}} \bra{\mathbf{s}^{\diamond}}\hat{A}\ket{\Psi}/ |\braket{\Psi} {\mathbf{s}^{\diamond}}|^2$ according to probabilities $Q(\mathbf{s}^{\diamond}) \equiv |\braket{\Psi} {\mathbf{s}^{\diamond}}|^2$,
\begin{equation}
 	\bra{\Psi} \hat{A} \ket{\Psi} = \sum_{\mathbf{s^{\diamond}}\in\mathcal{S}^{\diamond}} Q(\mathbf{s}^{\diamond}) A^{\diamond}(\mathbf{s}^{\diamond}).
\label{eq:average_dia}
\end{equation}
Similarly, we could generalize Eqs. \ref{eq:sampling}-\ref{eq:importance} and apply importance sampling. We note that in this case the variance $\sigma_{A^{\diamond}}^2$, defined by
\begin{eqnarray}
	\sigma_{A^{\diamond}}^2 &\equiv& \sum_{\mathbf{s}^{\diamond}} Q(\mathbf{s}^{\diamond}) |A^{\diamond}(\mathbf{s}^{\diamond})-\bar{A}|^2 \\
	&=& \sum_{\mathbf{s}^{\diamond}} Q(\mathbf{s}^{\diamond}) |A(\mathbf{s}^{\diamond})|^2-|\bar{A}|^2,
\end{eqnarray}	
might be smaller than the variance $\sigma_{\hat{A}}^2$ of operator $\hat{A}$ (Eq. \ref{eq:varianceAhat}), since a single incomplete sample $\mathbf{s}^{\diamond}$ corresponds to many complete samples $\mathbf{s}$. [For instance, in the example of Fig. \ref{fig:sampling}(c), the incomplete sample $\mathbf{s}^{\diamond} = (s_1,s_2,s_3)$ corresponds to all complete samples $\mathbf{s} =(s_1,s_2,s_3,s_4,s_5,s_6)$ that coincide with $\mathbf{s}^{\diamond}$ in the first three sites.] In other words, the statistical error might be reduced. This should not come as a surprise. After all, in the extreme case where no sampling at all is performed ($L^{\diamond} = 0$) but all indices are exactly contracted, there is no statistical error left.

\subsection{Markov chain Monte Carlo}

In Refs. \onlinecite{Schuch08, Sandvik07} the random configurations $\mathbf{s}$ were generated by means of a Markov chain process based on local updates. Given a stored configuration $\mathbf{s}$, let us denote $\mathbf{s}'_i$ a configuration obtained from $\mathbf{s}$ by replacing in site $i$ the value  $s_i$ with $s_i'$. Then, visiting the sites sequentially, $i=1,2, \cdots,L$, in what is known as a \textit{sweep}, a change on site $i$ is introduced according to the Metropolis probability
\begin{equation}
	P_{\mbox{\tiny change}} = \min[\frac{Q(\mathbf{s_i'})}{Q(\mathbf{s})},1].
\end{equation}
In this way, after one sweep a new configuration $\mathbf{s}'$ is obtained from $\mathbf{s}$, and by iteration a sequence of configurations
\begin{equation}
	\mathbf{s} \rightarrow \mathbf{s}' \rightarrow \mathbf{s}'' \rightarrow \cdots
	\label{eq:sequence}
\end{equation}
is produced. However, these configurations will in general be correlated. The number $\tau$ of sweeps required between configurations $\mathbf{s}$ and $\mathbf{s'}$ in order for them to be essentially independent to be independent is known as the autocorrelation time. Sweeping $\tau$ times between samples is necessary in order for the error $\epsilon_{A}(N)$ to scale as in Eq.~(\ref{eq:error}), since that expression for the error assumed the samples to be independent. (If only a single sweep mediates the samples, the statistical error in Eq.~(\ref{eq:error}) increases by a factor which scales as $\tau^{1/2}$ due to autocorrelations). In addition, the first sample $\mathbf{s}$ will be obtained after applying $\tau'$ sweeps to some random initial configuration. The equilibration time $\tau'$ is necessary in order to guarantee that the first sample is picked-up according to the correct probability distribution. The autocorrelation time $\tau$ and the equilibration time $\tau'$ are known to diverge with systems size $L$ for critical systems.

Large equilibration and autocorrelation times, e.g. near or at a critical point, increase the cost of simulations. This increase can be prevented if somehow independent configurations $\mathbf{s}$ can be directly generated according to probabilities $Q(\mathbf{s})$. In section \ref{sec:perfect} we show how this is possible for a specific class of tensor networks, namely unitary tensor networks, which are introduced next.

\section{Sampling of unitary tensor networks}
\label{sec:unitary}

\begin{figure}[t]
  \begin{centering}
    \includegraphics[width=8.5cm]{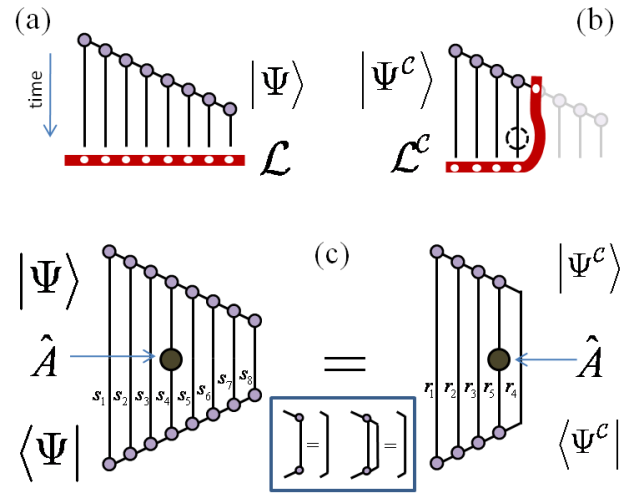}
    \caption{
    (Color online)
    Sampling in a unitary matrix product state (uMPS). (a) uMPS for a state $\ket{\Psi}$ of lattice $\mathcal{L}$. Notice the (fictitious) time direction, which provides each tensor with a sense of which indices are incoming and which are outgoing. (b) The past causal cone $\mathcal{C}$ of a local operator $\hat{A}$ acting on a single site of $\mathcal{L}$ (denoted by a discontinuous circle) defines an effective lattice $\mathcal{L}^{\mathcal{C}}$, which is found in state $\ket{\Psi^{\mathcal{C}}}$. Notice that the effective lattice $\mathcal{L}^{\mathcal{C}}$ is made of two types of sites, namely sites already present in the original lattice $\mathcal{L}$ and one site not present in $\mathcal{L}$, with $d$-dimensional and $\chi$-dimensional vector spaces, respectively.
    (c) Tensor networks representing  $\bra{\Psi}\hat{A}\ket{\Psi}$ and $\bra{\Psi^{\mathcal{C}} } \hat{A} \ket{\Psi^{\mathcal{C}}}$. The inset shows unitarity reductions [Eq.~(\ref{eq:unitary})] used to transform $\bra{\Psi}\hat{A}\ket{\Psi}$ into $\bra{\Psi^{\mathcal{C}} } \hat{A} \ket{\Psi^{\mathcal{C}}}$.
    \label{fig:causalMPS}}
  \end{centering}
\end{figure}

\begin{figure}[t]
  \begin{centering}
    \includegraphics[width=8.5cm]{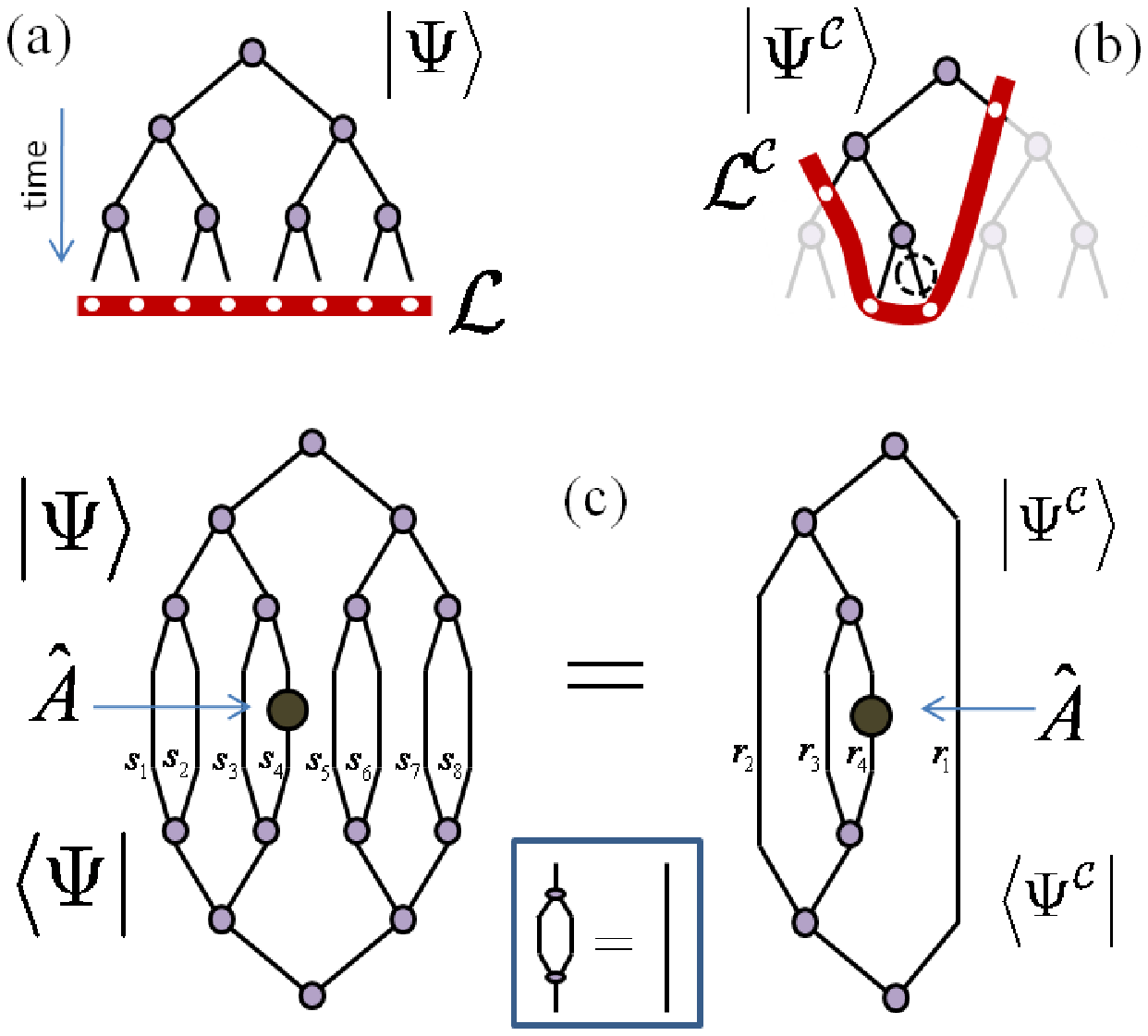}
    \caption{
    (Color online) Sampling in a unitary tree tensor network (uTTN).
    (a) uTTN for a state $\ket{\Psi}$ of lattice $\mathcal{L}$.
    (b) Effective lattice $\mathcal{L}^{\mathcal{C}}$.
    (c) Tensor networks for $\bra{\Psi}O\ket{\Psi}$ and $\bra{\Psi^{\mathcal{C}} } \mathcal{A} \ket{\Psi^{\mathcal{C}}}$. The inset shows a reduction due to the unitary constrain of tensors in the uTTN.
    \label{fig:causalTTN}}
  \end{centering}
\end{figure}

Let us specialize to the particular case of unitary tensor networks, namely tensor networks that are based on a unitary quantum circuit. Examples include the MERA and unitary versions of MPS (with open boundary conditions) and TTN, which we will refer as uMPS and uTTN \cite{unitary}.

Unitary tensor networks are special in that each tensor $u$ is constrained to be unitary/isometric. Figs. \ref{fig:causalMPS} and \ref{fig:causalTTN} exemplify the discussion for uMPS and uTTN respectively. Specifically, we first note that in one such tensor network there is a well-defined direction of time throughout, see e.g. Figs. \ref{fig:causalMPS}(a) and \ref{fig:causalTTN}(a). Each index of a tensor $u$ is either an incoming index (if time flows towards the tensor) or an outgoing index (if time flows away from the tensor). The constraint on $u$ can be expressed in the following way. Let us group all incoming indices of $u$ into a composite incoming index $\alpha$ and all outgoing indices of $u$ into a composite outgoing index $\beta$, so that tensor $u$ becomes a matrix $u_{\beta \alpha}$. Then the unitary/isometric constraint on $u$ reads
\begin{equation}
	\sum_{\beta} (u^{\dagger})_{\alpha \beta}u_{\beta \alpha'} = \delta_{\alpha\alpha'}.
	\label{eq:unitary}
\end{equation}
A direct implication of this property is that the tensor network corresponding to the expectation value $\bra{\Psi}\hat{A}\ket{\Psi}$ can be replaced with a simplified tensor network where the pairs of tensors $(u,u^{\dagger})$ outside the so-called past causal cone $\mathcal{C}$ of $\hat{A}$ have been removed, see Figs. \ref{fig:causalMPS}(c) and \ref{fig:causalTTN}(c). This new tensor network can be interpreted to represent the expectation value $\bra{\Psi^{\mathcal{C}} } \hat{A} \ket{\Psi^{\mathcal{C}}}$ of the local operator $\hat{A}$ on a state $\ket{\Psi^\mathcal{C}} \in \mathbb{V}_{\mathcal{L}^{\mathcal{C}}}$ of an effective lattice $\mathcal{L}^{\mathcal{C}}$ defined by the causal cone $\mathcal{C}$ of the operator $\hat{A}$, see Figs. \ref{fig:causalMPS}(b) and \ref{fig:causalTTN}(b), where by construction $\bra{\Psi} \hat{A} \ket{\Psi} = \bra{\Psi^{\mathcal{C}} } \hat{A} \ket{\Psi^{\mathcal{C}}}$. The effective lattice $\mathcal{L}^{\mathcal{C}}$ is made of $L^{\mathcal{C}}$ sites that can be of two types: those already contained in the original lattice $\mathcal{L}$, which are described by a $d$-dimensional vector space, and those which did not belong to $\mathcal{L}$, which are described by a $\chi$-dimensional vector space. We use $\mathbf{r} = (r_1,r_2,\cdots, r_{L^{\mathcal{C}}})$ to denote a configuration of the effective lattice $\mathcal{L}^{\mathcal{C}}$, and $\ket{\mathbf{r}} \equiv \ket{r_1}\otimes\ket{r_2}\otimes  \cdots \otimes \ket{r_{L^{\mathcal{C}}}}$ the corresponding product vector, where for some sites $r_i = 1, 2, \cdots, d$ and for some others $r_i = 1, 2, \cdots, \chi$. We denote $\mathcal{R}$ the set of all configurations $\mathbf{r}$.


\begin{figure}[t]
  \begin{centering}
    \includegraphics[width=8.5cm]{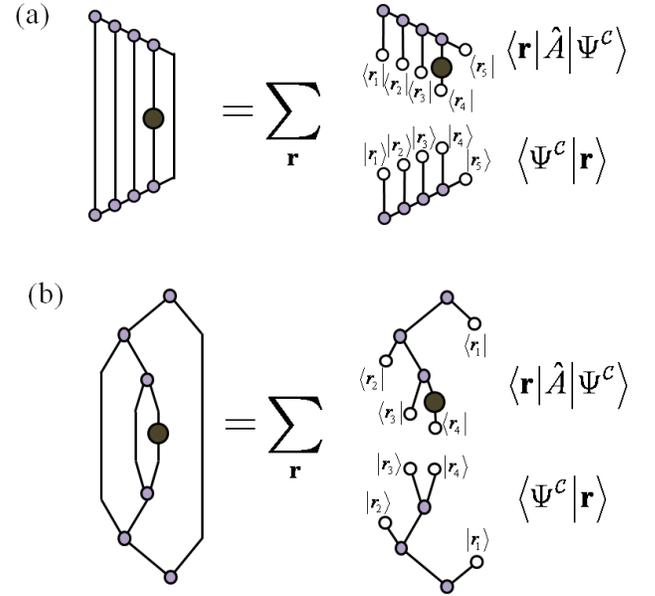}
    \caption{
    (Color online)
    Graphical representation of $\bra{\Psi^{\mathcal{C}}} \hat{A} \ket{\Psi^{\mathcal{C}}} = \sum_{\mathbf{r}\in\mathcal{R}} \braket{\Psi^{\mathcal{C}}}{\mathbf{r}}\bra{\mathbf{r}} \hat{A} \ket{\Psi^{\mathcal{C}}}$. In (a), the original state $\ket{\Psi}$ was represented with an uMPS, see Fig. \ref{fig:causalMPS}. In (b), the original state $\ket{\Psi}$ was represented with an uTTN, see Fig. \ref{fig:causalTTN}. However, in both cases the state $\ket{\Psi^{\mathcal{C}}}$ is represented by an uMPS that runs through the causal cone.
    \label{fig:sampleMPSandTTN}}
  \end{centering}
\end{figure}

The exact contraction of the tensor network corresponding to $\bra{\Psi^{\mathcal{C}} } \hat{A} \ket{\Psi^{\mathcal{C}}}$, may still be very expensive and again we might be interested in exploring the use of sampling to lower the computational cost. For that purpose, we repeat the discussion in section \ref{sec:background}. First we write the expectation value $\bra{\Psi} \hat{A} \ket{\Psi}$ as
\begin{equation}
\bra{\Psi} \hat{A} \ket{\Psi} = \sum_{\mathbf{r}\in\mathcal{R}} \braket{\Psi^{\mathcal{C}}}{\mathbf{r}}\bra{\mathbf{r}}\hat{A}\ket{\Psi^{\mathcal{C}}},
	\label{eq:expect2}
\end{equation}
see Fig. \ref{fig:sampleMPSandTTN} for uMPS and uTTN. Then we rewrite Eq.~(\ref{eq:expect2}) in terms of the estimator $A^{\mathcal{C}}(\mathbf{r})\equiv \bra{\mathbf{r}} \hat{A} \ket{\Psi^{\mathcal{C}}} / \braket{\mathbf{r}}{\Psi^{\mathcal{C}}}$ and probabilities $P(\mathbf{r}) \equiv |\braket{\mathbf{r}}{\Psi^{\mathcal{C}}}|^2$,
\begin{equation}
 	\bra{\Psi} \hat{A} \ket{\Psi} = \sum_{\mathbf{r}\in\mathcal{R}} P(\mathbf{r}) A^{\mathcal{C}}(\mathbf{r}).
\label{eq:average2}
\end{equation}
We can again limit the sum over configurations $\mathbf{r}$ to a subset $\tilde{\mathcal{R}}$ containing just $N$ configurations which, when chosen from $\mathcal{R}$ randomly according to the probabilities $P(\mathbf{r})$, results in
\begin{equation}
	\bra{\Psi} \hat{A} \ket{\Psi} \approx \frac{1}{N}\sum_{\mathbf{r}\in\tilde{\mathcal{R}}} A^{\mathcal{C}}(\mathbf{r}).
	\label{eq:importance2}
\end{equation}
The error in the approximation scales with $N$ as in Eq.~(\ref{eq:error}).

\section{perfect sampling}
\label{sec:perfect}

\begin{figure}[t]
  \begin{centering}
    \includegraphics[width=8.5cm]{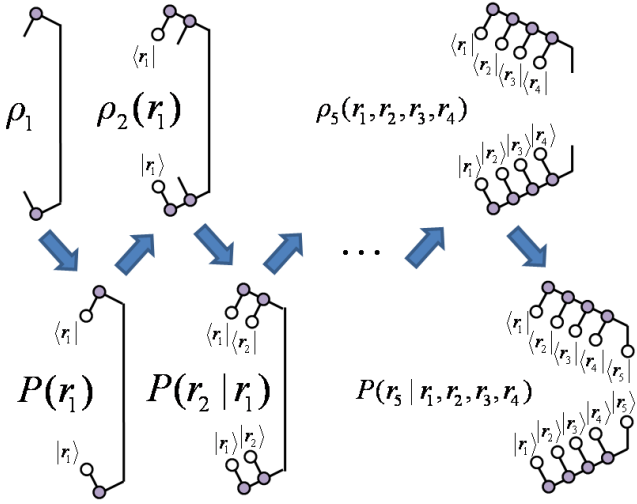}
    \caption{
    (Color online) Perfect sampling with a uMPS. The figure shows a sequence of the tensor networks corresponding (up to a proportionality constant) to $\rho_1$, $P(r_1)$, $\rho_2(r_2)$, $P(r_2|r_1)$, and so on, see Eqs.~(\ref{eq:P1}--\ref{eq:P123N}). Importantly, all these tensor networks can be contracted with a cost that scales as $O(\chi^2)$ with the bond dimension $\chi$, and are therefore computational less expensive than an exact contraction, which has cost $O(\chi^3)$.
    \label{fig:perfectMPS}}
  \end{centering}
\end{figure}

In this section we describe how to randomly draw configurations $\mathbf{r}$ according to probability $P(\mathbf{r})$ in a unitary tensor network. We refer to this scheme as perfect sampling because, in contrast with Markov chain Monte Carlo, the present scheme produces perfectly uncorrelated samples. We will also refer to this scheme as complete perfect sampling, to distinguish it from the incomplete perfect sampling scheme discussed in the next section, where sampling is performed only on a subset of sites.

\subsection{Algorithm}

Recall that as a quantum circuit, the tensor network is equipped with a notion of (fictitious) time. From now on we assume that the labeling of the sites in the effective lattice $\mathcal{L}^\mathcal{C}$ has been chosen so as to progress forward with respect to this notion of time. Thus, site $1$ corresponds to the earliest time, site $2$ corresponds to a later time, and so on, until site $L^{\mathcal{C}}$ corresponds to the latest time (when two sites correspond to the same time, e.g. sites $4$ and $5$ in Fig.~\ref{fig:sampleMPSandTTN}~(a), we order them arbitrarily).

Our \textit{perfect sampling} algorithm consists of sequentially computing a series of conditional single-site density matrices $\{\rho_1, \rho_2(r_1), \cdots \}$ and conditional single-site probabilities $\{P(r_1), P(r_2|r_1), \cdots\}$. First we compute the reduced density matrix $\rho_1$ for site $1$ exactly, i.e. without sampling,
\begin{equation}
	\rho_1 \equiv \tr_{2\cdots L^{\mathcal{C}}} \left\{ \ket{\Psi^{\mathcal {C}}} \bra{\Psi^{\mathcal{C}}} \right\}
	\label{eq:P1}
\end{equation}
from which we can compute the probabilities
\begin{equation}
	P(r_1) \equiv \bra{r_1} \rho_1 \ket{r_1}.
\end{equation}
We can then randomly choose a value for $r_1$ according to probability $P(r_1)$, and compute (exactly) the conditional reduced density matrix $\rho_2(r_1)$ for site $2$, which is obtained from the state $\braket{r_1}{\Psi^{\mathcal{C}}}$ of sites $2$ to $L^{\mathcal{C}}$,
\begin{equation}
	\rho_2(r_1) \equiv \frac{1}{P(r_1)}\tr_{3\cdots L^{\mathcal{C}}} \left\{ \braket{r_1}{\Psi^{\mathcal {C}}} \braket{\Psi^{\mathcal{C}}}{r_1}\right\}.
\end{equation}
Again, we can use the reduced density matrix to compute the conditional probabilities
\begin{equation}
	P(r_2|r_1) \equiv \bra{r_2} \rho_2(r_1) \ket{r_2},
\end{equation}
and we can therefore randomly select a value of $r_2$ according to probabilities $P(r_2|r_1)$.
Let us notice at this point that so far we have randomly chosen values for $r_1$ and $r_2$ according to the probability
\begin{equation}
	P(r_1,r_2) = P(r_1)P(r_2|r_1) = ||\braket{r_1,r_2}{\Psi^{\mathcal {C}}}||^2.
\end{equation}
We can now iterate the above process, that is, compute the conditional density matrix
\begin{equation}
	\rho_3(r_1,r_2) \equiv \frac{1}{P(r_1,r_2)} \tr_{4\cdots L^{\mathcal{C}}}  \left\{ \braket{r_1,r_2}{\Psi^{\mathcal {C}}} \braket{\Psi^{\mathcal{C}}}{r_1,r_2} \right \}
\end{equation}
and the conditional probabilities
\begin{equation}
	P(r_3|r_1,r_2) \equiv \bra{r_3} \rho_3(r_1,r_2) \ket{r_3},
\end{equation}
and so on for the rest of sites in the effective lattice $\mathcal{L}^{\mathcal{C}}$. In this way, and since
\begin{equation}
P(\mathbf{r}) =
P(r_1)P(r_2|r_1) \cdots  P(r_{L^{\mathcal{C}}}|r_1, r_2, \cdots, r_{L^{\mathcal{C}}-1}), \label{eq:P123N}
\end{equation}
we end up indeed randomly choosing a configuration $\mathbf{r} = (r_1,r_2, \cdots, r_{L^{\mathcal{C}}})$ with probability given precisely by $P(\mathbf{r})\equiv |\braket{\mathbf{r} }{\Psi^{\mathcal{C}}}|^2$.

Fig. \ref{fig:perfectMPS} illustrates the sequence of computations in the case of a one-site operator $\hat{A}$ specifically for a uMPS, assuming as in Figs. \ref{fig:causalMPS} and \ref{fig:sampleMPSandTTN}(a) that the operator $\hat{A}$ is supported on the fourth site of the original chain. This algorithm is similar to one used for thermal state sampling with MPS~\cite{White09} described in Ref.~\onlinecite{Stoudenmire10}. Analogous computations for a uTTN are very similar, since the causal cone of a single-site operator $\hat{A}$ is described also by a uMPS, see Fig. \ref{fig:causalTTN}(b). For the case of a MERA, more details on the implementation of Eqs.~(\ref{eq:P1}--\ref{eq:P123N}) can be found in Ref. \onlinecite{Ferris11}.

A key point is that, for unitary tensor networks such as uMPS, uTTN, and MERA, the computational cost of generating the above sequence of density matrices and probabilities often does not exceed (to leading order in $\chi$ and effective size $L^{\mathcal{C}}$) the cost of a single sweep in Markov chain Monte Carlo \cite{TTN2site}.

\begin{figure}[t!]
 \begin{centering}
  \includegraphics[width=4cm]{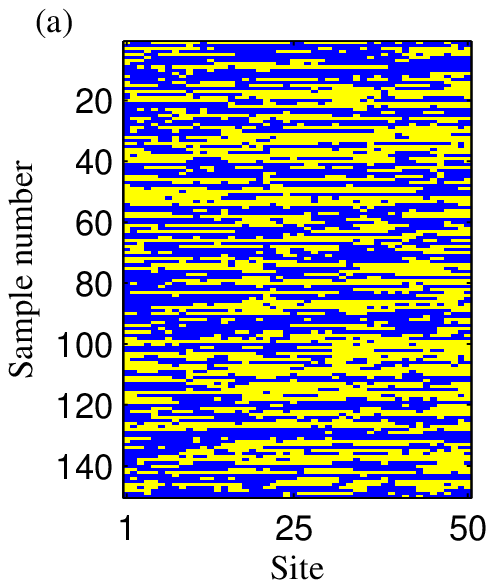}
  \includegraphics[width=4cm]{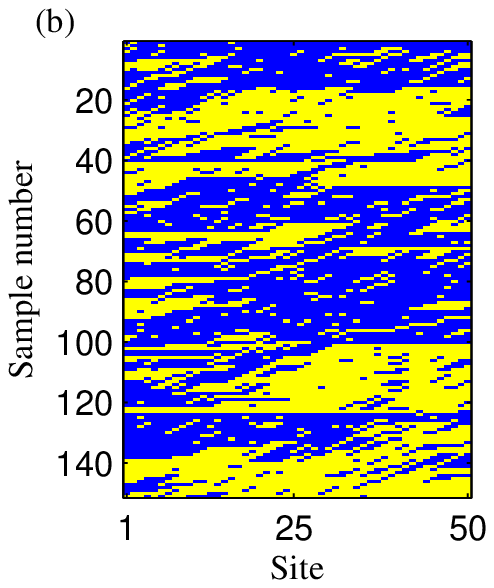}
  \includegraphics[width=4cm]{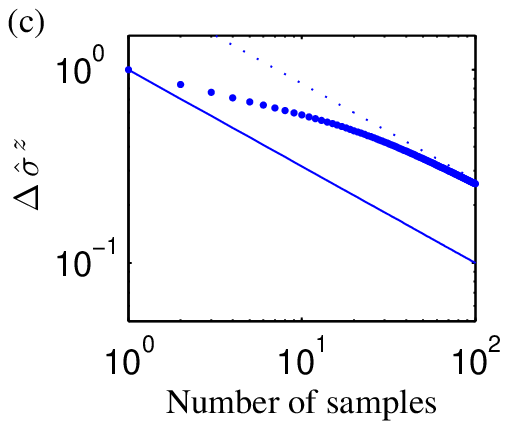}
  \includegraphics[width=4cm]{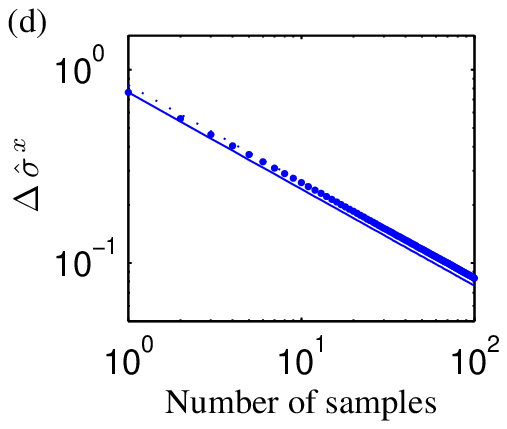}
  \includegraphics[width=4cm]{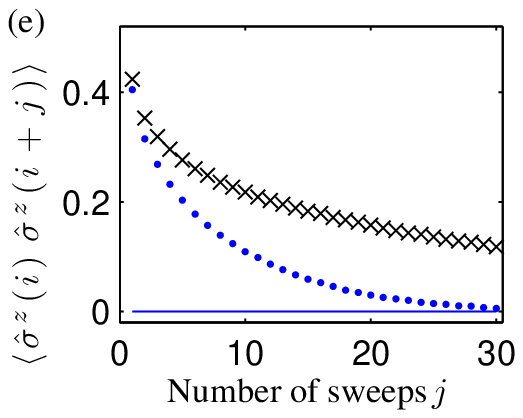}
  \includegraphics[width=4cm]{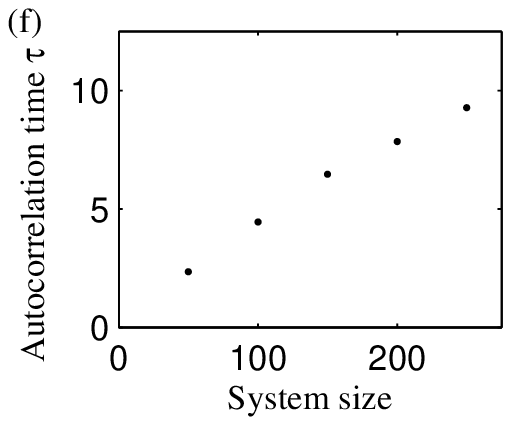}
  \caption{
  (Color online) Sampling of the ground state of the critical transverse Ising model in the $z$ basis. Comparison between configurations obtained using (a) the presented perfect sampling scheme and (b) a Markov chain scheme (single sweep) on 50 sites. Blue sites represent spin up and yellow for spin down. The correlations between configurations obtained using a Markov chain scheme are evidenced by the appearance of domains of well defined color that extend vertically. In (c) we have calculated the expected statistical error on the estimate of $\langle \hat{\sigma}^z \rangle$ for the perfect sampling (blue line) and Markov chain sampling (blue dots). While with perfect sampling the error decreases with the usual $N^{-1/2}$ factor, correlations between subsequent samples increase the error on the estimate in the Markov chain scheme. In (d) we plot the same for $\langle \hat{\sigma}^x \rangle$ by projecting all the spins into the $x$ basis. In this case the Markov scheme used utilizes a 2-site update so as to be compatible with the wave-function symmetry~\cite{twositeupdate}. In (e) we present the correlations on the centre site (in the $z$ basis) after $j$ Markov chain sweeps using $10^6$ samples for 50 sites (blue dots) and 250 sites (black crosses). In the perfect sampling scheme (blue line), there are no correlations between configurations. In (f) we plot the estimated autocorrelation time for different system sizes.
  \label{fig:comparison}}
 \end{centering}
\end{figure}

\subsection{Benchmark}

To illustrate the performance of the perfect sampling scheme and compare it to Markov chain Monte Carlo, we have considered a duly optimized uMPS for the ground state $\ket{\Psi}$ of the quantum Ising model with critical transverse magnetic field,
\begin{equation}
	\hat{H}_{\mbox{\tiny Ising}} \equiv -\sum_{\langle i, j \rangle} \hat{\sigma}^z_i \hat{\sigma}^z_j -\sum_i \hat{\sigma}^x_i,
	\label{eq:ising}
\end{equation}
on an open chain of $L$ spins.\cite{hamiltoniannote} The two sampling schemes are then used in order to compute the expectation value of local operators.

Fig. \ref{fig:comparison}(a) and (b) show a history of 150 configurations of a chain of $L=50$ spins obtained with perfect sampling and Markov chain Monte Carlo, respectively. The existence of correlations in the second case is manifest.

Fig. \ref{fig:comparison}(c) and (d) show the error in the expectation value $\bra{\Psi}\hat{\sigma}^z_{25} \ket{\Psi}$ and $\bra{\Psi}\hat{\sigma}^z_{25} \ket{\Psi}$ for the local operators $\sigma^z$ and $\sigma^x$ on site $25$, as a function of the number of samples $N$. In both cases, the effect of autocorrelations in Markov chain Monte Carlo results in an error larger than the error obtained with perfect sampling, which is given by Eq.~(\ref{eq:error}). The ratio between statistical errors, as given in terms of the autocorrelation time $\tau$ by $\sqrt{2\tau+1}$, is seen to depend on the choice of local operator -- this autocorrelation time is larger for $\bra{\Psi}\hat{\sigma}^z_{25} \ket{\Psi}$ than for $\bra{\Psi}\hat{\sigma}^x_{25} \ket{\Psi}$.

Finally, Fig. \ref{fig:comparison} (e) and (f) explore the autocorrelation time $\tau$ for $\hat{\sigma}^z$ as a function of the size $L$ of the spin chain. In particular, Fig. \ref{fig:comparison} (f) reveals that $\tau$ grows linearly in $L$. This means \cite{assumptions} that in order to achieve a fixed accuracy in $\bra{\Psi}\hat{\sigma}^z_{L/2} \ket{\Psi}$, the number of samples $N$ with Markov chain Monte Carlo has to grow linearly in $L$, whereas a constant number of samples is enough with perfect sampling.

It is important to stress, however, that the Markov chain Monte Carlo update scheme discussed here, based on single spin updates, is used as a reference only -- more sophisticated Markov chain Monte Carlo schemes, based e.g. on global spin updates, could lead to smaller autocorrelation times.

\section{Incomplete perfect sampling}
\label{sec:partial}

So far we have considered perfect sampling over the whole causal cone, that is, over the indices associated to all the sites of the effective lattice $\mathcal{L}^{\mathcal{C}}$. However, it is also possible to use an incomplte perfect sampling scheme, which combines perfect sampling over most of the sites of $\mathcal{L}^{\mathcal{C}}$ and an exact contraction over a small set of sites, without altering the scaling $O(\chi^q)$ of the cost of a single sample. Because we are sampling over fewer indices, we can expect a decrease in the statistical error with little change in the cost. In some cases the reduction in statistical uncertainty can be dramatic.

\begin{figure}[t]
  \begin{centering}
    \includegraphics[width=6cm]{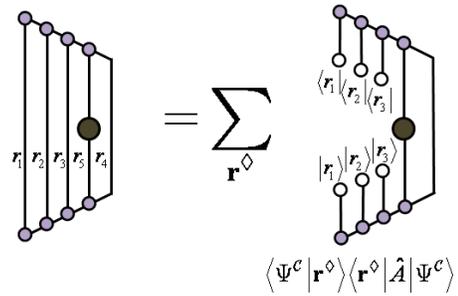}
    \caption{
    (Color online) Graphical representation of $\bra{\Psi^{\mathcal{C}}} \hat{A} \ket{\Psi^{\mathcal{C}}} = \sum_{\mathbf{r}^{\diamond} \in \mathcal{R}^{\diamond}} \braket{\Psi^{\mathcal{C}}}{\mathbf{r}}\bra{\mathbf{r}^{\diamond}} \hat{A} \ket{\Psi^{\mathcal{C}}}$ for a uMPS, to be compared with Fig. \ref{fig:sampleMPSandTTN}(a). Notice that sampling does not affect two of the indices, over which an exact contraction is still performed.
    \label{fig:partialMPS1}}
  \end{centering}
\end{figure}

\begin{figure}[t]
  \begin{centering}
    \includegraphics[width=7cm]{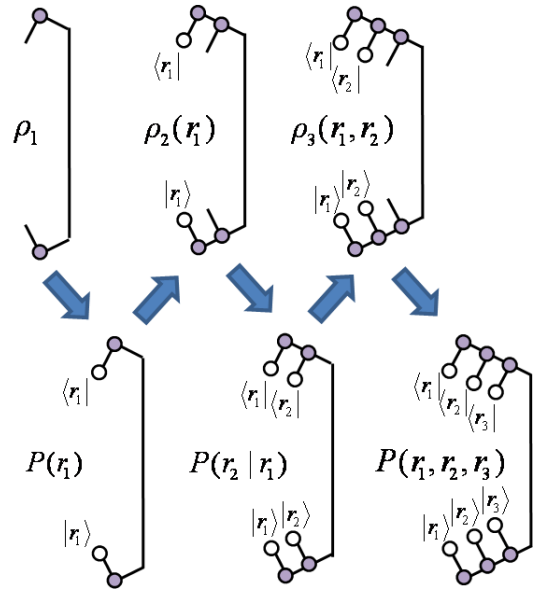}
    \caption{
    (Color online)
    Incomplete perfect sampling with a uMPS. The figure shows a complete sequence of the tensor networks corresponding (up to a proportionality constant) to $\rho_1$, $P(r_1)$, $\rho_2(r_2)$, $P(r_2|r_1)$, $\rho_3(r_1,r_2)$ and $P(r_1,r_2,r_3)$ necessary in order to generate a configuration $\mathbf{r}^{\diamond}=(r_1,r_2,r_3)$ with probability $P(\mathbf{r}^{\diamond}) = |\braket{\Psi^{\mathcal{C}}} {\mathbf{r}^{\diamond}}|^2$. Notice that the cost still scales as $O(\chi^2)$, as in the complete (perfect) sampling scheme.
    \label{fig:partialMPS2}}
  \end{centering}
\end{figure}

\subsection{Incomplete perfect sampling scheme}

The incomplete perfect sampling scheme is illustrated in Fig. \ref{fig:partialMPS1} for a uMPS. The first step is to rewrite the expectation value $\bra{\Psi^{\mathcal{C}}}\hat{A}\ket{\Psi^\mathcal{C}} = \bra{\Psi} \hat{A} \ket{\Psi}$ as
\begin{equation}
\bra{\Psi} \hat{A} \ket{\Psi} = \sum_{\mathbf{r^{\diamond}}\in\mathcal{R}^{\diamond}} \braket{\Psi^{\mathcal{C}}}{\mathbf{r}^{\diamond}}\bra{\mathbf{r}^{\diamond}}\hat{A}\ket{\Psi^{\mathcal{C}}},
	\label{eq:expect3}
\end{equation}
where $\mathcal{R}^{\diamond}$ is the set of incomplete configurations $\mathbf{r}^{\diamond} \equiv (r_1, r_2, \cdots ,r_{L^{\diamond}})$, where $L^{\diamond}$ is the number of sites over which sampling takes place, with $L^{\diamond} < L^{\mathcal{C}}$. For the case of the uMPS illustrated in Fig. \ref{fig:partialMPS1}, one can perform an exact contraction on two sites of $\mathcal{L}^{\mathcal{C}}$, namely the site on which the local operator $\hat{A}$ is supported and the effective, $\chi$-dimensional site corresponding to the bond index of the uMPS. Notice that now the term $\braket{\Psi^{\mathcal{C}}} {\mathbf{r}^{\diamond}} \bra{\mathbf{r}^{\diamond}}\hat{A}\ket{\Psi^{\mathcal{C}}}$ does not factorize into two terms, since $\braket{\mathbf{r}^{\diamond}}{\Psi^{\mathcal{C}}}$ and $\bra{\mathbf{r}^{\diamond}}\hat{A}\ket{\Psi^{\mathcal{C}}}$ are no longer complex numbers but $d\chi$-dimensional vectors.

We can still rewrite Eq.~(\ref{eq:expect3}) as a probabilitistic sum of an estimator $A^{\diamond}(\mathbf{r}^{\diamond}) \equiv  \braket{\Psi^{\mathcal{C}}} {\mathbf{r}^{\diamond}} \bra{\mathbf{r}^{\diamond}}\hat{A}\ket{\Psi^{\mathcal{C}}}/ |\braket{\Psi^{\mathcal{C}}} {\mathbf{r}^{\diamond}}|^2$ according to probabilities $P(\mathbf{r}^{\diamond}) \equiv |\braket{\Psi^{\mathcal{C}}} {\mathbf{r}^{\diamond}}|^2$,
\begin{equation}
 	\bra{\Psi} \hat{A} \ket{\Psi} = \sum_{\mathbf{r^{\diamond}}\in\mathcal{R}^{\diamond}} P(\mathbf{r}^{\diamond}) A^{\diamond}(\mathbf{r}^{\diamond}),
\label{eq:average3}
\end{equation}
limiting the sum over configurations $\mathbf{r}^{\diamond}$ to a subset $\tilde{\mathcal{R}}^{\diamond}$ containing just $N$ configurations, and use (perfect) importance sampling to obtain the estimate
\begin{equation}
	\bra{\Psi} \hat{A} \ket{\Psi} \approx \frac{1}{N}\sum_{\mathbf{r}^{\diamond }\in\tilde{\mathcal{R}}^{\diamond}} A^{\diamond}(\mathbf{r}^{\diamond}).
	\label{eq:importance3}
\end{equation}
An important difference between the incomplete perfect sampling scheme and the comnplete perfect sampling scheme of Eqs.~(\ref{eq:expect2}--\ref{eq:importance2}) is that the estimator $A^{\diamond}$, whose mean is $\bar{A^{\diamond}} = \bra{\Psi} \hat{A} \ket{\Psi}$ as indicated in Eq.~(\ref{eq:average3}), has a variance $\sigma^2_{A^{\diamond}}$,
\begin{eqnarray}
 \sigma^2_{A^{\diamond}} &\equiv& \sum_{\mathbf{r^{\diamond}}\in\mathcal{R}^{\diamond}} P(\mathbf{r}^{\diamond}) |A^{\diamond}(\mathbf{r}^{\diamond}) - \bar{A^{\diamond}}|^2 \\
 &=& \sum_{\mathbf{r^{\diamond}}\in\mathcal{R}^{\diamond}} P(\mathbf{r}^{\diamond}) |A^{\diamond}(\mathbf{r}^{\diamond})|^2 - |\bar{A^{\diamond}}|^2,
\end{eqnarray}
that is no longer necessarily equal to the variance $\sigma^2_{\hat{A}}$ of Eq.~(\ref{eq:varianceAhat}), but is instead upper bounded by it, $\sigma^2_{A^{\diamond}} \leq \sigma^2_{\hat{A}}$, see the Appendix. In other words, the error $\epsilon_{A^{\diamond}}(N)$ in the approximation of Eq.~(\ref{eq:importance3}), given by
\begin{equation}
	\epsilon_{A^{\diamond}}(N) \approx \sqrt{\frac{\sigma^{2}_{A^{\diamond}}}{N}},
\end{equation}
can be smaller than the error $\epsilon_{A}(N)$ of a complete sampling scheme.


\subsection{Algorithm}

We have implemented the incomplete perefect sampling scheme in conjunction with the complete perfect sampling scheme described in section \ref{sec:perfect}. We notice, however, that incomplete sampling can also be incorporated into Markov chain Monte Carlo.

As in section \ref{sec:perfect}, we proceed by constructing a sequence of conditional single-site reduced density matrices $\{\rho_1, \rho_2(r_1), \cdots \}$ and conditional probabilities $\{ P(r_1), P(r_2|r_1), \cdots \}$. However, in this occasion the sequence concludes at site $L^{\diamond}$, after which we can already evaluate the estimator $A^{\diamond}(\mathbf{r}^{\diamond})$. This is illustrated for the case of a uMPS in Fig. \ref{fig:partialMPS2}, which is to be compared with Fig. \ref{fig:perfectMPS}.

\subsection{Benchmark}

\begin{figure}[t]
 \begin{centering}
  \includegraphics[width=7.5cm]{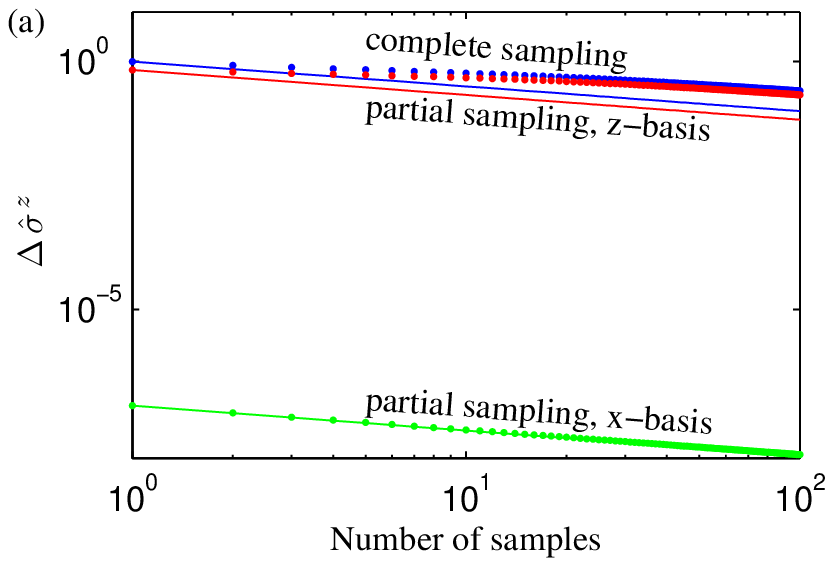}
  \includegraphics[width=7.5cm]{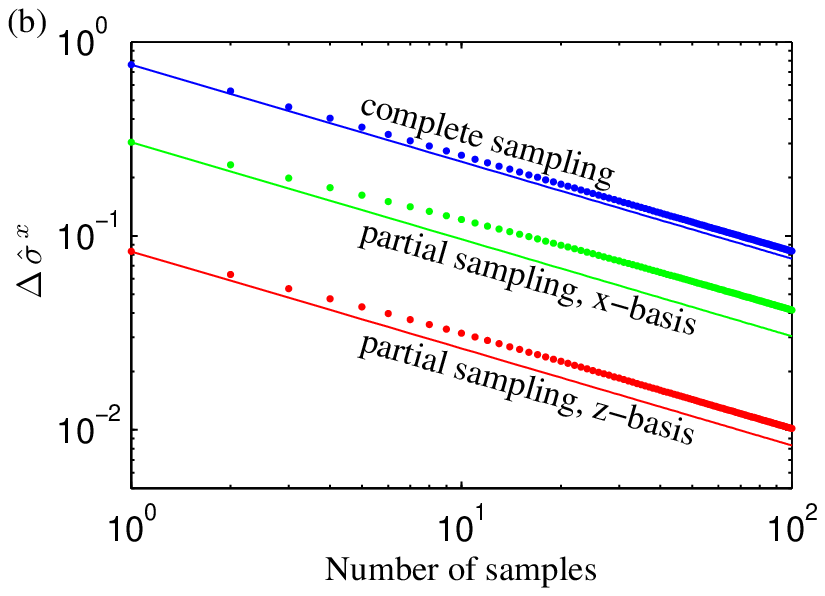}
  \caption{
  (Color online)
  Sampling errors with the incomplete perfect sampling scheme for a 50 site critical Ising chain, using both perfect sampling (continuous lines) and Markov chain Monte Carlo sampling (dots). (a) Sampling errors in the computation of $\bra{\Psi} \hat{\sigma}^z_{25} \ket{\Psi}$. With perfect sampling, errors in the incomplete perfect sampling scheme are upper-bounded by the errors in a complete sampling scheme, as proven in the Appendix. Interestingly, for estimates of $\langle \hat{\sigma}_z \rangle$ the incomplete perfect sampling scheme obtains an error $10^{-7}$ times smaller by measuring in the $x$ basis on sites $1,2,\cdots, L^{\diamond}$. (b) Sampling errors in the computation of $\bra{\Psi} \hat{\sigma}^x_{25} \ket{\Psi}$. Again, the errors with incomplete perfect sampling are smaller than those with complete perfect sampling, and depend on the choice of product basis.
  \label{fig:partialMPS3}}
 \end{centering}
\end{figure}

As in section \ref{sec:perfect}, we use sampling to compute the expectation value of local observables from a uMPS with $\chi = 30$ that has been previously optimized to approximate the ground state of the quantum Ising chain at criticality, Eq.~(\ref{eq:ising}). The exact structure that we sample can bee seen in Fig. \ref{fig:partialMPS1}. Figure~\ref{fig:partialMPS3} shows the sampling error, as a function of the number of samples $N$, in the computation of $\bra{\Psi} \hat{\sigma}^z_{25} \ket{\Psi}$ and $\bra{\Psi} \hat{\sigma}^x_{25} \ket{\Psi}$ in a chain of $L=50$ spins. The error is seen to depend on two factors. On the one hand, it depends on which operator ($\hat{\sigma}^z$ or $\hat{\sigma}^x$) is being measured, as it did in section \ref{sec:perfect}. In addition, now it also drastically depends on which product basis $\{ \ket{\mathbf{r}^{\diamond}} \}$ is used. In particular, we see that a very substantial reduction of sampling error, of seven orders of magnitude, is obtained by measuring on the $x$ basis while computing $\bra{\Psi} \hat{\sigma}^z_{25} \ket{\Psi}$. It should be noted that the two-site Markov chain update scheme used for the $x$-basis calculations,\cite{twositeupdate} although appears competitive, is more computationally demanding than the perfect sampling scheme and runs approximately 2--3 times slower.

\section{Computational costs}
\label{sec:costs}

For completeness, we include a brief summary of the computational costs incurred in extracting, from a given unitary tensor network, the expectation value of a local operator by using (i) exact contraction, (ii) Markov chain Monte Carlo and (iii) a perfect sampling scheme. For simplicity, we consider only one-site local operators. The scaling of the costs in the bond dimension $\chi$ is presented in Table \ref{table:cost}. We emphasize that in the sampling schemes, we only consider the cost of obtaining one sample. A fair comparison of costs with an exact contraction should also take into account the number of samples required in order to approximate the exact result with some pre-agreed accuracy.

\begin{table}[t]
 \begin{centering}
  \caption{The leading-order costs of contracting unitary tensor networks with and without sampling techniques, with the goal of estimating the expectation value of a one-site operator. For the MERA we have also included the cost calculating arbitrary (long-range) two-point correlators.\cite{Ferris11}
  \label{table:cost}}
  \begin{tabular}{|c|c|c|c|}
  \hline
  Tensor & Exact & Markov- & Perfect \\
  network & contraction & chain MC & sampling \\ \hline
  uMPS (open BC) & $\mathcal{O}(\chi^3)$ & $\mathcal{O}(\chi^2)$ & $\mathcal{O}(\chi^2)$ \\
  uTTN (binary) & $\mathcal{O}(\chi^4)$ & $\mathcal{O}(\chi^2)$ & $\mathcal{O}(\chi^3)$ \\
  MERA (1D binary) & $\mathcal{O}(\chi^9)$ & $\mathcal{O}(\chi^5)$ & $\mathcal{O}(\chi^5)$ \\
  $\rightarrow$ 2-point correlators & $\mathcal{O}(\chi^{12})$ & $\mathcal{O}(\chi^7)$ & $\mathcal{O}(\chi^8)$ \\
  \hline
  \end{tabular}
 \end{centering}
\end{table}

The table shows that for both a uMPS and the MERA, the cost of Markov chain Monte Carlo and perfect sampling scale with the same power. Instead, for the uTTN, the of Markov chain Monte Carlo is one power smaller than that of perfect sampling. [The same would happen with uMPS if the local dimension of each site was also $\chi$]. More significant speed-ups can be seen with the MERA, both for the computation of two-point correlators, and in systems in two dimensions (not in the Table), where sampling techniques to increase computational efficiency are required most. The authors present an in-depth analysis of perfect sampling with the MERA in Ref.~\onlinecite{Ferris11}.

A further remark is in order. The above analysis assumes that a tensor network has been provided in a unitary circuit form. In particular, the costs in Table \ref{table:cost} do not include operations such as converting a non-unitary version of the tensor network into its unitary form (typically through the QR-decomposition). In particular, the cost of QR-decompositions required to turn an MPS into a uMPS scales as $\mathcal{O}(\chi^3)$ -- that is, the same scaling as an exact contraction. What is then the practical interest in a perfect sampling scheme for a uMPS? On the one hand, the uMPS might conceivably have been generated through some procedure (e.g. along the lines of the algebraic Bethe Ansatz MPS constructions described in Ref. \onlinecite{algebraic}), with a cost $O(\chi^2)$ (notice that a uMPS tensor only contains $O(\chi^2)$ coefficients). In this case, the perfect sampling scheme would allow for a very efficient, approximate evaluation of expectation values without increasing this cost. On the other hand, although we have focused our analysis on the evaluation of local expectation values, more complex tasks involving a uMPS, such as the computation of entanglement entropy, can exploit the perfect sampling schemes presented in this paper at a cost significantly lower than that of an exact contraction (see e.g. Ref. \onlinecite{Cincio}).


\section{Conclusions}
\label{sec:conclusion}

We have explained how to perform Monte Carlo sampling on unitary tensor networks such as the MERA, uMPS and uTTN. In order to compute the expectation value $\bra{\Psi}\hat{A}\ket{\Psi}$ of a local operator $\hat{A}$, sampling is performed on the past causal cone $\mathcal{C}$ of operator $\hat{A}$. In addition, by exploiting the unitary character of the tensors, it is possible to directly sample configurations $\mathbf{r}$ of the causal cone according to their weight in the wave-function, resulting in uncorrelated samples and thus avoiding the equilibration and autocorrelation times of Markov chain Monte Carlo schemes. This last property makes the perfect sampling scheme particularly interesting to study critical systems.

In principle, one can also proceed as in Eqs.~(\ref{eq:P1}--\ref{eq:P123N}) for non-unitary tensor networks, e.g. PEPS, and obtain perfect sampling. However, in non-unitary tensor networks the cost of computing e.g. $\rho_1$ is already the same as that of computing the expectation value $\bra{\Psi}\hat{A}\ket{\Psi}$ without sampling. Therefore perfect sampling in non-unitary tensor networks seems to be of very limited interest.

Here we have only considered sampling in the context of computing expectation values. However, the same approach can also be applied in order to optimize the variational ansatz, as discussed in full detail in Ref. \onlinecite{Ferris11} for the MERA.

The authors thank Glen Evenbly for useful discussions. Support from the Australian Research
Council (FF0668731, DP0878830, DP1092513), the visitor programme at Perimeter Institute, NSERC and FQRNT is acknowledged.

\appendix
\section{Variance with complete and incomplete sampling}

Given a vector $\ket{\Psi}\in \mathbb{V}_{\mathcal{L}}$ and a local operator $\hat{A}$, the expectation value of $\hat{A}$ is given by $\bra{\Psi}\hat{A} \ket{\Psi}$ and its variance is
\begin{eqnarray}
\sigma_{\hat{A}}^2	&\equiv&  \bra{\Psi} \left( |\hat{A}-\bra{\Psi}\hat{A}\ket{\Psi}|^2 \right) \ket{\Psi} \\
&=& \bra{\Psi} \left( |\hat{A}|^2 \right) \ket{\Psi}-|\bra{\Psi} \hat{A} \ket{\Psi}|^2.
\label{eq:varianceAhatApp}
\end{eqnarray}

\subsection{Mean and variance with complete sampling}

Consider the complex random variable $(A(\mathbf{s}),P(\mathbf{s}))$, where $A(\mathbf{s})$ is the estimator
\begin{equation}
A(\mathbf{s}) \equiv \frac{ \braket{\Psi}{\mathbf{s}} \bra{\mathbf{s}} \hat{A} \ket{\Psi}}{ \braket{\Psi}{\mathbf{s}}\braket{\mathbf{s}}{\Psi}} = \frac{\bra{\mathbf{s}} \hat{A} \ket{\Psi}}{\braket{\mathbf{s}}{\Psi}}
\end{equation}
and $Q(\mathbf{s})$ is the probability
\begin{equation}
	Q(\mathbf{s}) \equiv \braket{\Psi}{\mathbf{s}}\braket{\mathbf{s}}{\Psi}.
\end{equation}
Here $\{\ket{\mathbf{s}}\}$ denotes an orthonormal basis in the vector space $\mathbb{V}_{\mathcal{L}}$. Notice that $\sum_{\mathbf{s}} \proj{\mathbf{s}}$ is a resolution of the identity in $\mathbb{V}_{\mathcal{L}}$ and therefore $\sum_{\mathbf{s}} Q(\mathbf{s}) = \braket{\Psi}{\Psi} = 1$.

The mean $\bar{A}$ is given by the expectation value $\bra{\Psi}\hat{A}\ket{\Psi}$,
\begin{eqnarray}
	\bar{A} &\equiv& \sum_{\mathbf{s}} Q(\mathbf{s})A(\mathbf{s})
	= \sum_{\mathbf{s}} \braket{\Psi}{\mathbf{s}}\braket{\mathbf{s}}{\Psi} \frac{ \braket{\Psi}{\mathbf{s}} \bra{\mathbf{s}} \hat{A} \ket{\Psi}}{ \braket{\Psi}{\mathbf{s}}\braket{\mathbf{s}}{\Psi}} \nonumber \\
	&=& \sum_{\mathbf{s}}\braket{\Psi}{\mathbf{s}} \bra{\mathbf{s}} \hat{A} \ket{\Psi} =
	\bra{\Psi} \hat{A} \ket{\Psi}.
\end{eqnarray}
In turn, its variance $\sigma^2_{A}$,
\begin{eqnarray}
	\sigma_{A}^2 &\equiv& \sum_{\mathbf{s}} Q(\mathbf{s}) |A(\mathbf{s})-\bar{A}|^2 \\
	&=& \sum_{\mathbf{s}} Q(\mathbf{s}) |A(\mathbf{s})|^2-|\bar{A}|^2,
\end{eqnarray}
equals the variance $\sigma^2_{\hat{A}}$ of operator $\hat{A}$, as can be seen from
\begin{eqnarray}
	 &&\sum_{\mathbf{s}} Q(\mathbf{s})|A(\mathbf{s})|^2
	= \sum_{\mathbf{s}} \braket{\Psi}{\mathbf{s}}\braket{\mathbf{s}}{\Psi} \frac{ \bra{\Psi} \hat{A}^{\dagger} \ket{\mathbf{s}}\bra{\mathbf{s}} \hat{A} \ket{\Psi}}{ \braket{\Psi}{\mathbf{s}}\braket{\mathbf{s}}{\Psi}} \nonumber \\
&&	= \sum_{\mathbf{s}} \bra{\Psi} \hat{A}^{\dagger} \ket{\mathbf{s}}\bra{\mathbf{s}} \hat{A} \ket{\Psi} = \bra{\Psi} \left(|\hat{A}|^2 \right)\ket{\Psi}.
\end{eqnarray}

\subsection{Mean and variance with incomplete sampling}

Consider now a new complex random variable $(A(\mathbf{s}), Q(\mathbf{s}))$, where $A(\mathbf{s})$ is the estimator
\begin{equation}
A(\mathbf{s}) \equiv \frac{ \bra{\Psi} \pi(\mathbf{s}) \hat{A} \ket{\Psi}}{ \bra{\Psi}\pi(\mathbf{s})\ket{\Psi}}
\end{equation}
and $Q(\mathbf{s})$ is the probability
\begin{equation}
	Q(\mathbf{s}) \equiv \bra{\Psi} \pi(\mathbf{s}) \ket{\Psi}.
\end{equation}
Here $\{\pi(\mathbf{s})\}$ denotes a complete set of projectors on the vector space $\mathbb{V}_{\mathcal{L}}$, that is $\pi(\mathbf{s})^2 = \pi(\mathbf{s})$, and $\sum_{\mathbf{s}} \pi(\mathbf{s})$ is a resolution of the identity in $\mathbb{V}_{\mathcal{L}}$, so that $\sum_{\mathbf{s}} Q(\mathbf{s}) = \braket{\Psi}{\Psi} = 1$. Notice that if all the projectors $\pi(\mathbf{s})$ have rank one, then we recover the situation analyzed in the previous subsection. Notice also that this more general setting includes the case addressed in Sect. \ref{sec:partial} in the context of incomplete sampling.

The mean $\bar{A}$ is again given by the expectation value $\bra{\Psi}\hat{A}\ket{\Psi}$,
\begin{eqnarray}
	\bar{A} &\equiv& \sum_{\mathbf{s}} Q(\mathbf{s})A(\mathbf{s})
	= \sum_{\mathbf{s}} \bra{\Psi}\pi(\mathbf{s})\ket{\Psi} \frac{ \bra{\Psi} \pi(\mathbf{s}) \hat{A} \ket{\Psi}}{ \bra{\Psi}\pi(\mathbf{s})\ket{\Psi}} \nonumber \\
	&=& \sum_{\mathbf{s}}\bra{\Psi} \pi(\mathbf{s}) \hat{A} \ket{\Psi} =
	\bra{\Psi} \hat{A} \ket{\Psi}.
\end{eqnarray}
However, this time the variance $\sigma^2_{A}$ is only upper bounded by the variance $\sigma^2_{\hat{A}}$ of operator $\hat{A}$. This follows from,
\begin{eqnarray}
&&\sum_{\mathbf{s}} Q(\mathbf{s})|A(\mathbf{s})|^2 \nonumber \\
&&	= \sum_{\mathbf{s}}  \bra{\Psi} \pi(\mathbf{s}) \ket{\Psi}
\frac{ \bra{\Psi} \hat{A}^{\dagger} \pi(\mathbf{s})  \ket{\Psi}}{ \bra{\Psi}\pi(\mathbf{s})\ket{\Psi}} \frac{ \bra{\Psi} \pi(\mathbf{s}) \hat{A} \ket{\Psi}}{ \bra{\Psi}\pi(\mathbf{s})\ket{\Psi}} \nonumber \\
&&= \sum_{\mathbf{s}}
 \frac{ \bra{\Psi} \hat{A}^{\dagger} \pi(\mathbf{s})  \ket{\Psi}\bra{\Psi} \pi(\mathbf{s}) \hat{A} \ket{\Psi}}{ \bra{\Psi}\pi(\mathbf{s})\ket{\Psi}} \nonumber \\
&&\leq \sum_{\mathbf{s}} \bra{\Psi} \hat{A}^{\dagger} \pi(\mathbf{s}) \hat{A} \ket{\Psi} = \bra{\Psi} \left(|\hat{A}|^2 \right)\ket{\Psi}.
\end{eqnarray}

Here, the inequality follows from $\braket{x}{y}\braket{y}{x} \leq \braket{x}{x}\braket{y}{y}$ with the identifications $\ket{x} \equiv \pi(\mathbf{s}) \hat{A}\ket{\Psi}$ and $\ket{y}\equiv \pi(\mathbf{s})\ket{\Psi}$.


\begin{thebibliography}{99}
\bibitem{reduction} Exploitation of space symmetries (e.g. translation invariance and scale invariance) can significantly reduce the computational cost of simulations from $O(L)$ to $O(\log L)$ or even to a constant (independent of $L$), allowing to reach the thermodynamic limit.
\bibitem{unitary} From an MPS (TTN) for a state $\ket{\Psi}$, with given bond dimension $\chi$, one can always use the gauge freedom in these tensor networks to obtain a unitary MPS (respectively TTN) for the same state $\ket{\Psi}$ and with the same bond dimension $\chi$, by writing the tensor network in its canonical form \cite{Vidal03, Shi06}.
\bibitem{TTN2site} We have found some tensor networks for which perfect sampling costs one more power of $\chi$ than a Markov chain sweep, see Ref.~\onlinecite{Ferris11}. Further, for local operators supported on two or more sites, perfect sampling may incur a slightly larger cost than that of a single sweep of Markov chain Monte Carlo. For example, this occurs with a TTN, where perfect sampling of a two-site operator costs $\chi^4$ instead of $\chi^3$, but not with MPS or MERA.
\bibitem{twositeupdate} The transverse Ising model contains a $\mathbb{Z}_2$ symmetry as the operator $\prod_i \hat{X}_i$ commutes with the Hamiltonian. For a wave-function chosen from one of the $\pm 1$ sectors of this operator, after making a projective measurement of all the spins in the $x$-basis, one always gets an even (odd) number number of spin downs (actually, lefts). The overlap after flipping a single spin of such a configuration is always zero, and thus a two-site Markov chain update scheme that can preserve parity is required.
\bibitem{hamiltoniannote} Our Hamiltonian is actually defined in terms of bond operators, leading to effectively half the magnetic field at the ends of the chain. This change does not affect any of the critical properties of the system.
\bibitem{assumptions} We assume that the variance $\bra{\Psi} (\hat{\sigma}^z_{L/2})^2 \ket{\Psi} - (\bra{\Psi} \hat{\sigma}^z_{L/2} \ket{\Psi})^2$, which is upper bounded by 1, is essentially constant as a function of the system size $L$.
\end{thebibliography}
\end{document}